\documentclass[]{aa}
\usepackage{amsmath}
\usepackage{mathrsfs}
\usepackage{graphicx}
\usepackage{natbib}
\bibpunct{(}{)}{;}{a}{}{,}

\def\suzaku{\textsl{Suzaku}\xspace}
\def\inte{\textsl{INTEGRAL}\xspace}

\newcommand{\xmm}{\textsl{XMM-Newton}\xspace}
\newcommand{\pn}{\text{EPIC-pn}\xspace}
\newcommand{\mos}{\text{EPIC-MOS}\xspace}
\newcommand{\rgs}{\text{RGS}\xspace}

\def\exo{EXO~2030+375\xspace}
\def\b12{E_{\rm 12}}

\begin{document}

\title{Glancing through the accretion column of \exo}

\author{
	Carlo Ferrigno\inst{1}
        \and 
        Patryk Pjanka\inst{2,3}
        \and
        Enrico Bozzo\inst{1}
        \and
        Dmitry Klochkov\inst{4}
        \and
        Lorenzo Ducci\inst{1,4}
        \and  
         Andrzej A. Zdziarski\inst{3}
}

\authorrunning{C. Ferrigno et al.}
\titlerunning{Glancing through an accretion column}
   \offprints{C. Ferrigno}

\institute{ISDC, Department of astronomy, University of Geneva, chemin d'\'Ecogia, 16 CH-1290 Versoix, Switzerland\\
	\email{carlo.ferrigno@unige.ch}
	\and
	Department of Astrophysical Sciences, Princeton University, 4 Ivy Lane, NJ 08544 Princeton, USA
         \and
    	Centrum Astronomiczne im.\ M. Kopernika, Bartycka 18, PL-00-716 Warszawa, Poland
         \and
         Institut f\"ur Astronomie und Astrophysik, Kepler Center for Astro and Particle Physics, Eberhard Karls Universit\"at, Sand 1, 72076 T\"ubingen, Germany
        }

\date{Received ---; accepted ---}

\abstract
	{The current generation of X-ray instruments is progressively revealing more and more details about  
	the complex magnetic field topology and the geometry of the accretion flows in highly magnetized accretion powered pulsars. } 
	{We took advantage of the large collecting area and good timing capabilities of the EPIC cameras on-board \xmm\ to 
	investigate the accretion geometry onto the magnetized neutron star hosted in the high mass X-ray binary 
	\exo\ during the rise of a source Type-I outburst in 2014.} 
         {We carried out a timing and spectral analysis of the \xmm\ observation as function of the neutron star spin phase. We used  
         a phenomenological spectral continuum model comprising the required 
         fluorescence emission lines. Two neutral absorption components are present:
         one covering fully the source and one only partially. The same analysis was also 
         carried out on two \suzaku\ observations of the source performed during outbursts in 2007 and 2012,
         to search for possible spectral variations at different luminosities. 
        } 
        {The \xmm\ data caught the source at an X-ray luminosity of $2\times10^{36}$\,erg\,s$^{-1}$ and revealed the presence of a narrow dip-like 
        feature in its pulse profile that was never reported before. The width of this feature corresponds to about one hundredth of 
        the neutron star spin period. From the results of the phase-resolved spectral analysis we suggest that this feature can be ascribed 
        to the self-obscuration of the accretion stream passing in front of the observer line of sight. 
        We inferred from the \suzaku\ observation carried out in 2007 that the self-obscuration of the accretion 
        stream might produce a significantly wider feature in the neutron star pulsed profile at higher luminosities   
        ($\gtrsim$$2\times10^{37}$\,erg\,s$^{-1}$).}  
	{This discovery allowed us to derive 
	additional constraints on the physical properties of the accretion flow in this object at relatively small distances 
	from the neutron star surface. The presence of such a narrow dip-like feature in the pulse profile 
	is so far unique among all known high mass X-ray binaries hosting 
	strongly magnetized neutron stars.} 

   \keywords{X-rays: binaries, stars: neutron, pulsars: individual: \exo }

\maketitle

\section{Introduction}
\label{sec:intro}

\exo\ is a prototypical high mass Be X-ray binary (BeXRB), comprising a neutron star (NS) 
and a Be companion.
X-ray outbursts are generally produced when the orbit of the pulsar intercepts 
the companion's decretion disk. Be/X-ray binaries typically show two types of
outbursts: (1) giant outbursts (type II), which lasts tens of days and 
are characterized by high luminosities and high spin-up rates (i.e., a significant increase
in pulse frequency), 
and (2) normal outbursts (type I), which are characterized
by lower luminosities and less pronounced spin-up rates (if any). 
Type I outbursts are known to occur (almost) regularly 
at each periastron passage in some systems \citep{stella1986, bildsten1997}. 
During these events, the material lost by the Be star is first focused toward the NS as a consequence of its    
strong gravitational field, and then funneled by its intense magnetic field ($B\sim10^{12-13}$\,G) down to the magnetic poles, 
where one or more accretion columns are formed.    

The bulk of the continuum X-ray emission from BeXRBs is produced within 
the accretion columns due to the Compton scattering 
of seed thermal photons from the hot spot on the NS surface or by  
bremsstrahlung processes occurring along the column \citep[][and references therein]{becker2007}. 
The spectral energy distribution of these sources is expected to show a 
remarkable dependence on the spin phase due to the changes in the viewing angle of the observer and the angular 
dependence of the Compton scattering cross section in a strong magnetic field \citep{meszaros1985a,meszaros1988b}. 
Iron fluorescence lines corresponding to different ionization levels of these heavy ions 
are commonly observed in BeXRBs and ascribed 
to the illumination of the accreting material at different 
distances from the NS by the intense 
X-ray radiation.  

The broad band spectra of BeXRBs are usually described by different phenomenological models, 
the most widely used one being an absorbed power-law modified at high energy by an exponential cut-off. 
Depending on the statistics of the data and the energy coverage, it proved necessary in several cases 
to complement these relatively simple spectral models with the additions of broad Gaussian components 
\citep[see, e.g.,][]{klochkov2007, suchy2008} and/or partial covering absorbers \citep{Naik2011,Naik2012,Naik2014}.
Furthermore, the cyclotron resonant scattering of electrons in the high magnetic field of the NS is known to 
produce characteristic absorption lines that have been observed in several of these systems and included in the 
spectral fits by using either Gaussian or Lorentzian profiles  
\citep[see, e.g.,][for a recent review]{walter15}. Cyclotron Resonant Scattering Features (CRSFs)
can be used to infer the NS surface magnetic field 
strength as the centroid energy of the fundamental line is at   
$E_{\rm cyc} \simeq 11.6\,B_{12}\times (1+z)^{-1}$\,keV, where $B_{12}$ 
is the NS magnetic field strength in units of $10^{12}$\,G
and $z$ the redshift of the scattering medium.

\exo hosts a 42\,s pulsar  
discovered with \textsl{EXOSAT} during a giant type-II outburst in 1985 \citep{parmar1989b}. 
The compact object orbits a B0\,Ve star \citep[][and references therein]{coe1988}
every 46 days \citep{wilson2005,wilson2008}. 
The estimated distance to the source is 7.1\,kpc \citep{wilson2002}. 
Type-I outbursts reaching peak fluxes of about 100\,mCrab (15-50\,keV) have been regularly detected 
from \exo at virtually all periastron passages since 1991 \citep[the outburst peak usually occurs about 
$\sim7$\,d after the periastron passage; see, e.g.,][]{wilson2005}. In the period spanning 
from 1992 to 1994, the type-I outbursts have been brighter than average and the NS showed a remarkable spin-up. 
From 1994 to 2002, the outbursts showed somewhat lower peak luminosities (a factor of few) and the 
pulsar displayed a clear spin-down trend. A new re-brightening period followed until  
June 2006, when \exo underwent its second observed giant type-II outburst. After a number of 
binary orbits characterized by a higher persistent luminosity than average and type-I outburst achieving a peak flux of 
200-300\,mCrab (15-50\,keV), the source returned back to its normal behavior. 

During type-I and type-II outbursts, the broad-band spectrum of \exo\ can be reasonably well described by using  
an absorbed ($N_\mathrm{H}\simeq10^{22}\,\mathrm{cm}^{-2}$) power-law with a high-energy
exponential roll-over. Contrasting results have been published concerning the presence of possible 
CRSFs in the X-ray emission from the source. \citet{reig1999} reported on the possible 
detection of such a feature with a centroid energy of $\sim36$\,keV, while \citet{wilson2008} 
found evidence of a CRSF at $\sim11$\,keV. 
\citet{klochkov2007} showed, however, that the data used by \citet{wilson2008} could also be reasonably well characterized without the 
cyclotron line by including in the fit a broad Gaussian emission feature at $\sim15$\,keV  
\citep[as observed in other high mass X-ray binaries; see, e.g., the discussion in][and references therein]{ferrigno2009}.  
\citet{klochkov2008} also reported on the detection of a CRSF
at $\sim63$\,keV in spin-resolved spectra extracted during the peak of the 2006 giant outburst. 
This could correspond to a higher harmonics of the previously suggested cyclotron line at $\sim36$\,keV. 

The pulse profile of \exo is known to be strongly dependent on the X-ray luminosity. At the peak of the outbursts 
its characteristic shape is usually interpreted in terms of a fan beam-like emission, while it become more reminiscent 
of what is expected in the case of a pencil beam emission during the decay of the outburst \citep{parmar1989b,klochkov2008}. 
A similar interpretation was also suggested by the detailed study carried out by \citet{sasaki2010} with a pulse decomposition method. 

In this paper, we report on the first \xmm\ observation of \exo performed during the rise of a type-I outburst in 2014. 
The large collecting area and good timing resolution of the EPIC-pn camera on-board \xmm\ allowed us to extract the 
source pulse profiles with more than 100 phase bins, revealing a peculiar sharp and deep feature never detected before 
(Sect.~\ref{sec:timing}). Following our spectral results (Sect.~\ref{sec:spectral}), we suggest 
that this feature is caused by the obscuration effect of the accretion stream 
passing in front of the observer line of sight to the source (Sect.~\ref{sec:phase}). 
The implications of our results are discussed in Sect.~\ref{sec:discussion} and summarized in Sect.~\ref{sec:conclusion}.

\section{Observations and data analysis}
\label{sec:observations}

We first analyze the \xmm\ observation of \exo\ which caught the source during the rise to the 
peak of a type-I outburst in 2014. The results of this observations are then compared with those obtained with \suzaku\ 
during type-I X-ray outbursts occurred in 2007 and 2012. A log of all observations is provided in Table~\ref{tab:observations}.
\begin{table*}
\caption{Log of all observations used in this paper.}
\begin{center}
 \begin{tabular}{ l c c c c c c }
\hline
\hline
Orbital phase     & Start Time [UT] & Stop Time [UT] & \multicolumn{3}{c}{Exposures [ks]} & $L_X$\tablefootmark{a} [erg/s] \\
             &                   &                   &{\small \pn}&{\small\mos1}&{\small\rgs} & \\ 

\hline
 0.022--0.030 &  2014-05-29 23:58 & 2014-05-30 8:30 &  30.2  & 31.7 &  32.8 & $2.1\times10^{36}$ \\
\hline
    &      &                   & {\small XIS}  & \multicolumn{2}{c}{\small HXD} &  \\ 
\hline
  0.027--0.068\tablefootmark{b}     & 2012-05-23 20:12  & 2012-05-25 18:57  & 77.9 &\multicolumn{2}{c}{ 67.7} & $2.7\times10^{36}$  \\        
  0.132--0.159\tablefootmark{c}     & 2007-05-14 20:37  & 2007-05-16 03:45  & 28.6 &\multicolumn{2}{c}{ 50.0} & $1.9\times10^{37}$ \\          
\hline
\end{tabular}
\tablefoot{
\tablefoottext{a}{In the 1--10 keV band and assuming a distance of 7.1 kpc.}\\
\tablefoottext{b}{Data set analyzed also in \citet{Naik2014}.}\\
\tablefoottext{c}{Data set analyzed also in \citet{Naik2012}.}\\
}
\label{tab:observations}
\end{center}
\end{table*}

\subsection{XMM-Newton}

During the \xmm\ observation of \exo\ in 2014 the \pn was operated in timing mode, while the MOS1 was in small window and 
the MOS2 in full frame. 
We followed standard data pipeline reduction procedures (\texttt{epchain}, \texttt{emchain}, \texttt{rgsproc}) by using the SAS v.13. 
No episodes of enhanced solar activity were revealed in the data, and thus we retained the full exposure available for all EPIC 
cameras. We extracted the \pn spectra and light curves of the source (background) from the CCD columns 34-43 (3-10). 
The average source count rate measured by the \pn was of 31 cts/s,  thus pile-up was not an issue for these data. The MOS 
data suffered instead a significant pile-up. To correct for this issue, we removed a progressively larger central part of the 
MOS point spread function for the source spectral extraction until a good match was obtained with the measured flux and spectral 
shape of the \pn data. Data from the MOS1 could be relatively well corrected by extracting the source spectra 
from an annular region with an inner radius of 150\,pixels (1.125\,arcmin) and an external radius of 1750\,pixels 
(1.46\,arcmin). 
The MOS1 background spectrum was 
extracted by using an equivalently large region from a different chip not contaminated by the source X-ray emission.  
The pile-up of the MOS2 turned out to be too sever to attempt any correction, and we thus discarded these data for further analysis. 
All EPIC spectra of the source were optimally rebinned using the prescription in paragraph 5.4 of \citet{Kaastra2016}.
The RGS spectra of the first dispersion order were characterized by a low S/N and were thus combined with the tool \texttt{rgscombine}. 
The spectra were then rebinned with a minimum number of 20 photons per energy bin. 
We did not make use of the second order spectra due to the significantly lower S/N. 

A first look to the source EPIC and RGS spectra revealed that they are heavily absorbed below 1\,keV.
As reported in a number of other published papers in the literature, we noticed that the 
\pn data in timing mode suffered redistribution issues below 1.7\,keV with such a large absorption column density. 
Significant discrepancies between the residuals in the \pn, MOS1, and the RGS appeared below 
this energy for any fit we attempted. For the final scientific analysis, we thus limited our fits to the energy range 
1.7-11\,keV for the \pn, 1-10\,keV for the MOS1, and 1-2.1\,keV for the two RGSs. We also excluded from the fits 
the \pn data in the energy range 2.19-2.39\,keV due to the instrumental residuals related to the gold edge 
in the effective area. This issue is linked to small uncertainties in the energy calibration 
for the fast modes coupled with the large number of detected counts \citep[see, e.g.,][and references therein]{ferrigno2014}.

\subsection{\suzaku}
 
We used the tools included in HEASoft v.6.14 and the latest calibration files available in CALDB for the 
HXD (13.09.2011) and the XIS (01.07.2014) to perform all \suzaku\ data analysis. The raw data from the two instruments  
were reprocessed using \texttt{aepipeline} separately for the XIS, the PIN, and the GSO detectors.  
The XIS data suffered a significant pile-up due to the brightness of the source and we used 
\texttt{pile\_estimate.sl}\footnote{\url{http://space.mit.edu/ASC/software/suzaku/pest.html}} to evaluate 
the annular extraction region required to filter out all events affected by a pile-up fraction $\gtrsim 5\%$. 
With this method, we reduced the data pile-up fraction from an original $\sim 8\%$ to $\sim 2\%$. 
The source and background spectra were derived from the cleaned and corrected event files using \texttt{xselect} and  
identical extraction annular regions. 
Spectra were grouped as described in the previous paragraph.
We generated the Redistribution Matrix Functions (RMFs) and Auxiliary Response 
Functions (ARFs) with \texttt{xisrmfgen} and \texttt{xissimarfgen}, respectively. The non X-ray background files (NXBs) 
were generated using \texttt{xisnxbgen} and combined with previously extracted backgrounds with \texttt{mathpha} 
(taking into account the proper scaling for the effective area). No NXB could be generated for the XIS1 detector due to the 
lack of data in the CALDB. Following the standard recommendations to avoid uncertainties in the effective area and response matrix 
of the XIS detectors, we considered only the energy range 1--10\,keV for the XIS0 and XIS3, and the energy range 1--8\,keV for 
the XIS1. The energy ranges 1.72--1.88\,keV, 2.19--2.37\,keV, and 1.72--2.37\,keV were excluded for the XIS0, XIS3, and XIS1, respectively,  
due to the known Au and Si calibration features \citep{Nowak2011, Kuehnel2013}.
The background models and response files for the HXD, PIN, and GSO are supplied by the mission team 
\footnote{\url{http://www.astro.isas.jaxa.jp/suzaku/analysis/hxd/}}. We used the ``tuned'' version of these files for the PIN. 
The PIN and GSO spectra were extracted using \texttt{hxdpinxbpi} and \texttt{hxdgsoxbpi}.

\section{Timing analysis}
\label{sec:timing}

To carry out a proper timing analysis of the \xmm data, we first converted the arrival time of each photon detected by the \pn and MOS1  
to the solar system barycenter using the known optical position of the source \citep{yCat2003}. The newly obtained arrival times  
were then converted to the system of the binary line of the nodes using the first solution for the orbital ephemeris of the source 
published in Table 2 of \citet{wilson2002}. The spin period of the source in the \xmm\ data was then determined  
at 41.2858(8)\,s (at 1$\sigma$ c.l.) using an epoch folding technique. This value is compatible with the FERMI/GBM measurement at the same epoch.\footnote{The Fermi/GBM measurements are available at the URL \url{http://gammaray.nsstc.nasa.gov/gbm/science/pulsars/lightcurves/exo2030.html }.}

\begin{figure}[]
\begin{center}
\resizebox{\hsize}{!}{\includegraphics[angle=0]{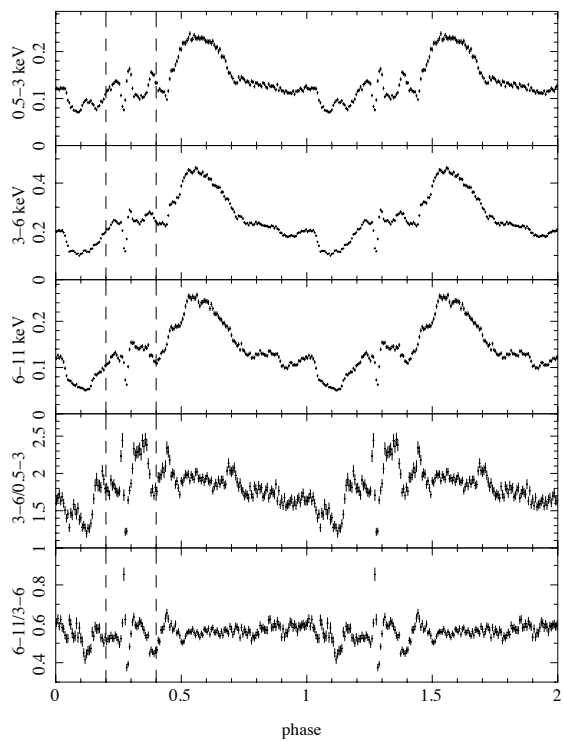}}
\caption{Pulse profiles and hardness ratios extracted from the \pn data in the 0.5-3\,keV, 3--6\,keV, and 6--11\,keV energy bands. 
The two lowermost panels shows the ratio between the counts in the hard and soft energy bands. Vertical dotted lines highlight the zoom region of Fig.~\ref{fig:pulses_zoom}.}
\label{fig:pulses}
\end{center}
\end{figure}

\begin{figure}[]
\begin{center}
\resizebox{\hsize}{!}{\includegraphics[angle=0]{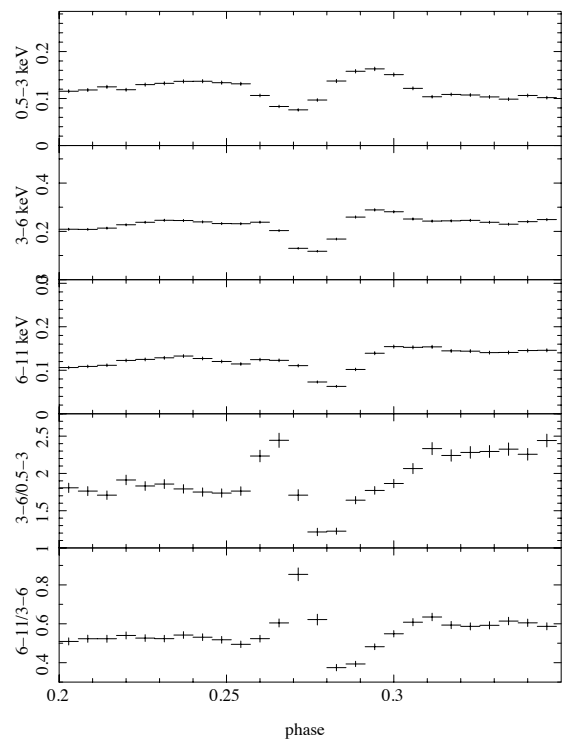}}
\caption{Same as Fig.~\ref{fig:pulses}, but zoomed around the pronounced narrow feature at phase $\sim$0.27.}
\label{fig:pulses_zoom}
\end{center}
\end{figure}

Pulse profiles were extracted from the \pn\ and MOS1 data with 175 phase bins in three energy bands,
chosen 
to equally split the number of photons.
Their shape (see Fig.~\ref{fig:pulses}) is very similar to that  
observed by \suzaku\ in 2012 \citep[see][and our Fig.~\ref{fig:phase_resolved_suzaku_12}]{Naik2012}, when the source was at a 
comparable luminosity. Remarkable swings in the hardness ratio are present 
at phases $\lesssim$0.5. 
Additionally, we noticed in the \pn\ pulse profile the presence 
of a sharp V-like feature at phase 0.27 (right before the main pulse profile peak) that was never reported before.
A zoom of the data closer to the phase of this feature is presented in Fig.~\ref{fig:pulses_zoom}. Its 
profile displays a noticeable energy dependence, comprising an increased absorption in the soft X-rays 
at phases $\lesssim$0.27 and an enhanced re-emission at harder X-rays. The peak of the re-emission occurs 
at later phases for harder X-rays. 

For the timing analysis of the 2007 and 2012 \suzaku\ data we followed the same  
procedures as in \citet{Naik2012} and \citet{Naik2014}, respectively. The derived pulse profiles are shown in 
Fig.~\ref{fig:phase_resolved_suzaku_12} and \ref{fig:phase_resolved_suzaku_07} and are fully in agreement 
with those published previously.  

\section{Phase averaged spectral analysis}
\label{sec:spectral}

A number of sophisticated spectral models have been developed to describe the X-ray energy distribution 
of highly magnetized accreting X-ray pulsars \citep[e.g.,][]{becker2007,farinelli2012,farinelli2016}. They include a fairly 
detailed description of the physics of the NS accretion column, where the bulk of the X-ray emission is 
produced (see Sect.~\ref{sec:intro}). 
In this paper, we focus on the analysis of the sharp feature highlighted in Sect.~\ref{sec:timing}, 
as well as on any possible spectral change related to it, and thus we did not attempt to fit the \xmm\ data 
of \exo\ with these models because the energy range of the EPIC cameras is too limited to provide significant 
constraints to all their free parameters. 
Yet, these models are not designed to perform spin phase-resolved spectral analysis. 
We thus opted for a phenomenological description, comprising a 
power law modified by an exponential cutoff at high energy:
\begin{equation}
N(E)=\left\{
\begin{array}{ll}
E^{-\Gamma}  & \mbox{for } E\le E_\mathrm{C}\,,\\
E^{-\Gamma} \exp\left(-\frac{E-E_\mathrm{C}}{E_\mathrm{F}}\right) & \mbox{for } E>E_\mathrm{C}. 
\end{array}
\right.
\label{eq:continuum}
\end{equation}
In the equation above, $E_\mathrm{C}$ and $E_\mathrm{F}$ are the cutoff and folding energies, respectively.
We used this model to fit for both the phase-averaged 
and phase-resolved \xmm\ and \suzaku\ spectra. 

We included in the spectral model
a first absorption component (\texttt{tbnew\_feo})\footnote{\url{http://pulsar.sternwarte.uni-erlangen.de/wilms/research/tbabs/}}, representing the 
summed contribution of the Galactic interstellar medium along the line of sight to the source and  
the local wind material from the massive companion surrounding the NS (100\% coverage). We also introduced a second, partially covering
neutral absorption component (\texttt{tbnew\_pcf})
to take into account the X-ray extinction due to material closer to the NS and/or within its accretion column. 
The parameters of this absorption component are the column density of the 
absorber $N_\mathrm{H,partial}$ and its covering fraction $f$ (which represents the fraction of the total radiation 
from the source that is affected by the presence of the absorbing material).
We checked that the introduction of a redshift for the neutral absorber material, expected due to the strong influence 
of the NS gravitational field on the radiation emerging close to the surface of the compact object, could not 
be reasonably constrained by our dataset.
We note that this absorber is expected to be at least partly ionized, due to the 
relatively high X-ray luminosity of the source. However, when we attempted to use a partially covering ionized absorber model 
(\texttt{warmabs}\footnote{\url{http://heasarc.gsfc.nasa.gov/xstar/docs/html/node102.html}}), 
the fit revealed that the ionization 
parameter could only be constrained to be  $\log_{10} \xi < -3$ at 90\% c.l, i.e., no significant ionization could be measured.
All absorption components used the element abundances reported 
by \citet{Wilms2000} and cross sections taken from \citet{Verner1996}.

We completed the spectral model with two Gaussian emission lines to take into account the neutral K$\alpha$ iron line at 6.4\,keV, 
as well as the the K$\beta$ line at 7.05~keV. 
Additional lines corresponding to higher ionization stages (\ion{Fe}{xvi}) of the iron ions 
were detected only in the 2007 \suzaku\ data, when the source was significantly brighter.
In this observation, we also detected the \ion{Si}{xiv} at 2.5\,keV and the \ion{S}{xv} and 3.2\,keV, similarly to \citet{Naik2012}, while the 
\ion{S}{xiii} line at 2\,keV lies in the boundary of our energy selection windows and cannot be effectively constrained.
The width of all Gaussian 
lines was fixed to zero in those fits where no significant broadening could be measures and was left free to very in all other cases 
(see Table~\ref{tab:phase_averaged}).
To account for calibration uncertainties, we have added a 1\% systematics during the fits. Due to known \suzaku 
calibration issues, we left free to vary the power-law photon index in the spectrum extracted from the back-illuminated CCD (XIS1) 
with respect to those measured from the front-illuminated chips and from the PIN instrument. A minor difference was obtained 
($\sim$0.05), but this was enough to cause a significant improvement in the resulting $\chi^2$. In the 2007 observation, we obtained 
$\Delta\chi^2$=97 (from 462 to 461 d.o.f.). when we allowed different power-law slopes. For the 2012 data, the improvement was 
$\Delta\chi^2$=109 passing from 383 to 382 d.o.f..

A comparison between the results obtained with the above fits to the \xmm and \suzaku data collected in 2012 reveals that there is a 
good agreement with the properties of the partial absorber, while the continuum parameters show some significant deviations.
We are currently unable to determine whether the properties of the source are 
intrinsically different or if this can be due to inter-calibration issues of the instruments. We verified that the different energy ranges covered 
by the two facilities are not affecting this outcome, as neglecting the PIN data in \suzaku only resulted in a slightly worse constrained 
cutoff and folding energies. We comment more on this point in Sect.~\ref{sec:discussion}.
 
All best fit parameters are reported in Table~\ref{tab:phase_averaged}. Cross-calibration constants were introduced in all fits 
and fixed to unity for the \pn and the XIS0, as we used these  
instruments as references for the \xmm\ and \suzaku\ observations, respectively. All spectral fits have been performed by 
using \texttt{XSPEC} v12.8.1g \citep{XSPEC}.

\begin{table*}
\caption{Spectral parameters obtained from the best fits to the \xmm\ and \suzaku\ phase averaged data.}
\begin{center}
 \begin{tabular}{ l r@{}l r@{}l  r@{}l}
\hline
\hline
&\multicolumn{2}{c}{\xmm}&\multicolumn{2}{c}{\suzaku 2012} &\multicolumn{2}{c}{\suzaku 2007}\\
\hline
$E_\mathrm{Fe K\alpha}$ [keV] & 6.50 & $\pm$0.01 &  6.41 & $\pm$0.01 & 6.38 & $\pm$0.02\\ 
\smallskip
$N_\mathrm{Fe K\alpha}$ [ph\,s$^{-1}$\,cm$^{-2}$] & (1.0 &$\pm$0.1)$\times10^{-4}$ & (1.8 &$\pm$0.2)$\times10^{-4}$ & (6 &$\pm$1)$\times10^{-4}$\\ 
\smallskip
$N_\mathrm{Fe K\beta}$\tablefootmark{a} [ph\,s$^{-1}$\,cm$^{-2}$]  & $<$2&$\times10^{-5}$ & (4 &$\pm$1)$\times10^{-5}$  & -&-\\
\smallskip
 $E_\mathrm{\ion{Fe}{xxvi}}$ [keV] & -&-& -&- &  6.63 & $\pm$0.02 \\
\smallskip
$\sigma_\mathrm{\ion{Fe}{xxvi}}$[keV] & -&-& -&- & 0.07 & $\pm$0.03\\
\smallskip
$N_\mathrm{\ion{Fe}{xxvi}}$ [ph\,s$^{-1}$\,cm$^{-2}$] & -&- & -&- &  (8&$\pm$2)$\times10^{-4}$\\
\smallskip
 $E_\mathrm{\ion{Si}{xiv}}$ [keV] & -&-& -&- &  2.53 & $^{+0.10}_{-0.04}$\\
\smallskip
$N_\mathrm{\ion{Si}{xiv}}$ [ph\,s$^{-1}$\,cm$^{-2}$] & -&- & -&- &  (5&$\pm$2)$\times10^{-4}$\\
\smallskip
 $E_\mathrm{\ion{S}{xv}}$ [keV] & -&-& -&- &   3.19 & $\pm$0.04 \\
\smallskip
$\sigma_\mathrm{\ion{S}{xv}}$ [keV] & -&-& -&- &   0.09 & $^{+0.04}_{-0.03}$  \\
\smallskip
$N_\mathrm{\ion{S}{xv}}$ [ph\,s$^{-1}$\,cm$^{-2}$] & -&- & -&- &  (8&$\pm$3)$\times10^{-4}$\\
\smallskip
$N_\mathrm{H}$ [$10^{22}$cm$^{-2}$] & 2.84 & $\pm$0.07 &  3.13 & $\pm$0.05  & 2.97 & $\pm$0.02 \\ 
\smallskip
$N_\mathrm{H,pc}$ [$10^{22}$cm$^{-2}$] & 8.6 & $\pm$0.4 & 8.1 & $\pm$0.4 &  124 & $\pm$18\\ 
\smallskip
$f$ &0.59 & $\pm$0.01  & 0.58 & $\pm$0.01  & 0.24 & $\pm$0.03 \\
\smallskip
$\Gamma$  & 1.17 & $\pm$0.02  & 1.33 & $\pm$0.03 & 1.31 & $\pm$0.01 \\
\smallskip
$\Gamma_\mathrm{XIS1}$  & -- &-- &  1.38 & $\pm$0.03 &  1.27 & $\pm$0.01\\
\smallskip
$E_\mathrm{C}$ [keV] &  7.9 & $\pm$0.2 &  7.0 & $\pm$0.3  &  6.8 & $^{+0.2}_{-0.1}$ \\ 
\smallskip
$E_\mathrm{F}$ [keV] &  18 & $\pm2$ &  26 & $\pm$1  &  20.9 & $\pm$0.4 \\ 
\smallskip
Flux$_\mathrm{PL,(0.5-10 keV)}$\tablefootmark{b}  &5.72 & $\pm$0.07 & 7.53&$\pm0.13$ & 59&$^{+3}_{-2}$\\
\smallskip
$\chi^2_\mathrm{red}$(d.o.f.)  & 1.158\tablefootmark{c}&(456) & 1.091\tablefootmark{c}&(382) & 1.106\tablefootmark{c}&(461)  \\
\hline
$C_\mathrm{MOS1}$ &  1.157 & $\pm$0.005 & -&- & -&- \\ 
$C_\mathrm{RGS}$ &  0.27\tablefootmark{d} & $\pm$0.01 &  -&- & -&-  \\
$C_\mathrm{XIS1}$ & -&-&   1.045 & $\pm$0.005& 0.995 & $\pm$0.003  \\
$C_\mathrm{XIS3}$ & -&- & 1.054 & $\pm$0.004 & 0.968 & $\pm$0.002 \\
$C_\mathrm{PIN/GSO}$ & -&- & 1.18 & $\pm$0.03 & 1.08 & $\pm$0.04 \\
\hline
\end{tabular}
\tablefoot{
\tablefoottext{a}{The centroid energy is fixed to be 0.65 keV higher than that of the K$\alpha$ line. The widths of the Gaussian 
lines are fixed to zero when not explicitly reported.}
\tablefoottext{b}{The flux of the power-law component without exponential cutoff and absorption is a fit parameter 
expressed in units of $10^{-10}$erg\,s$^{-1}$\,cm$^{-2}$.}
\tablefoottext{c}{A 0.5\% (1\%) systematic error is added in quadrature for \xmm (\suzaku) data.}
\tablefoottext{d}{This small ($<<1$) inter-calibration constant is due to a technical feature of XSPEC 
and reflects the smaller energy range of the RGS response, as compared to the EPIC cameras.}
}
\label{tab:phase_averaged}
\end{center}
\end{table*}

\begin{figure}[]
\begin{center}
\resizebox{\hsize}{!}{\includegraphics[angle=0]{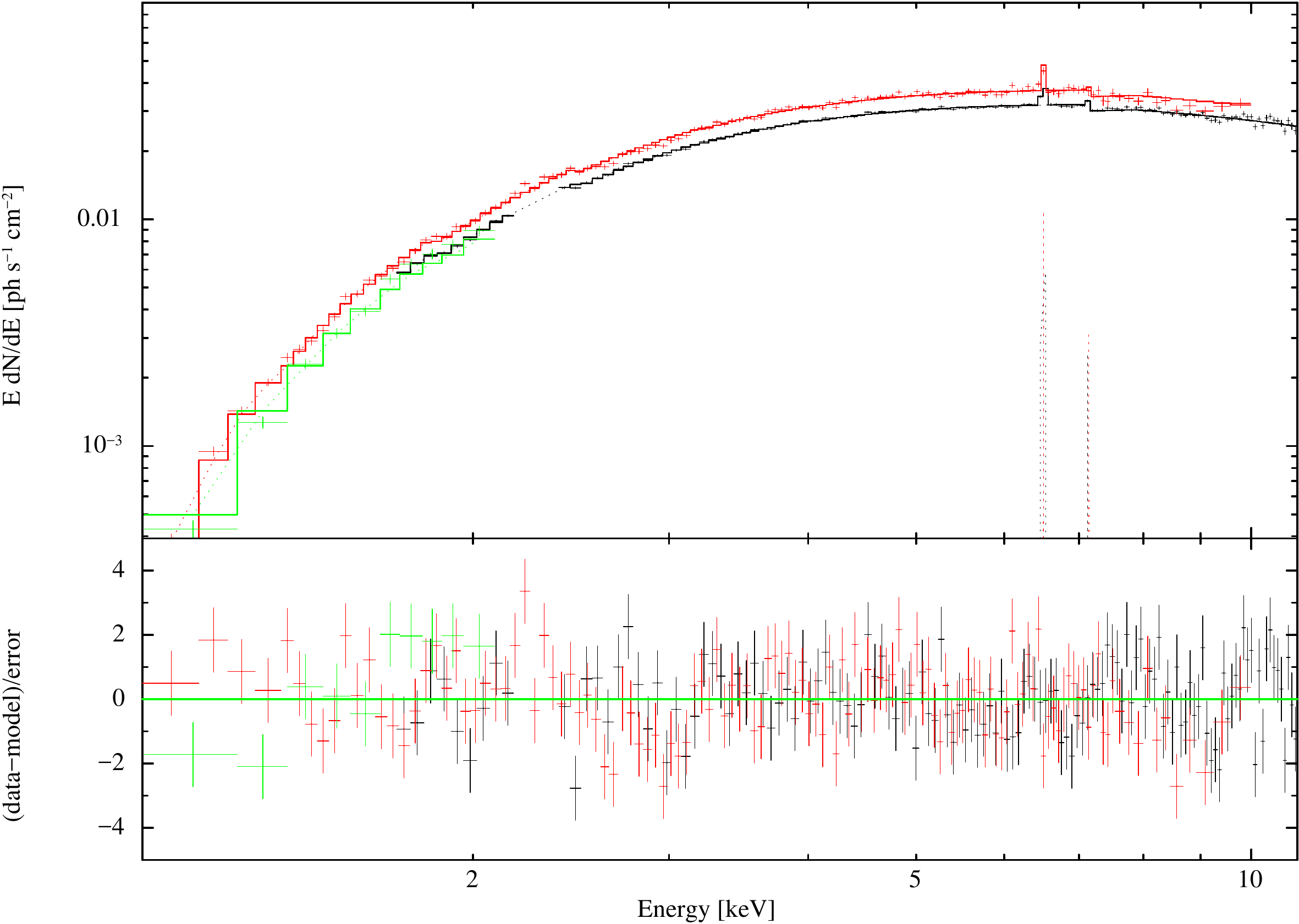}}
\caption{Phase-averaged unfolded energy spectrum of the \xmm observation carried out in 2014. 
All best fit parameters are reported 
in Table~\ref{tab:phase_averaged}. \pn (black), \mos (red), and \rgs (green) data have been 
rebinned for plotting purposes.}
\label{fig:phase_averaged}
\end{center}
\end{figure}

\begin{figure}[]
\begin{center}
\resizebox{\hsize}{!}{\includegraphics[angle=0]{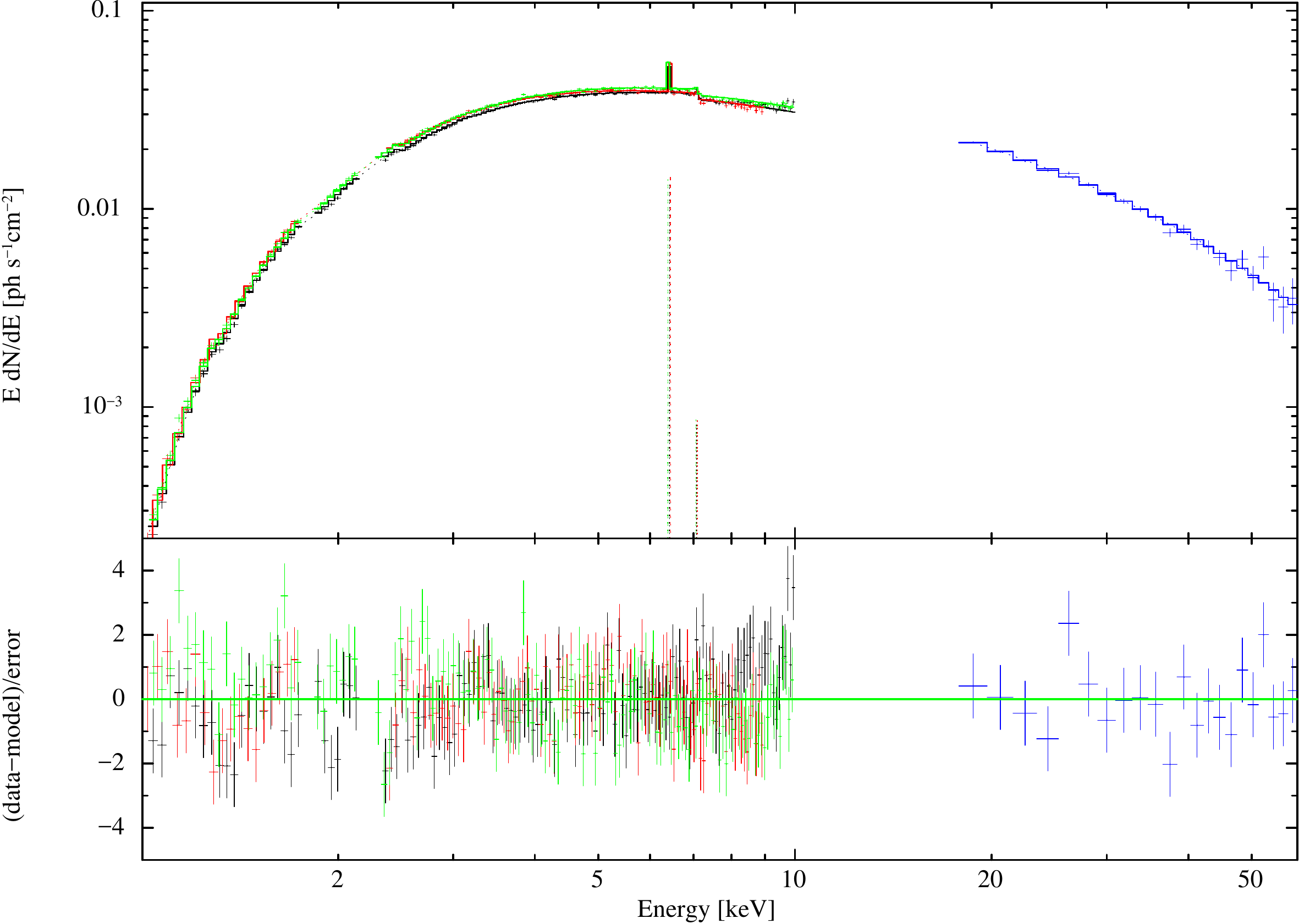}}
\caption{Same as Fig.~\ref{fig:phase_averaged} but for the \suzaku\ data collected in 2012. 
XIS0 (black), XIS1 (red), XIS3 (green), and HXD-PIN data (blue) have been 
rebinned for plotting purposes.}
\label{fig:phase_averaged_2012}
\end{center}
\end{figure}

\begin{figure}[]
\begin{center}
\resizebox{\hsize}{!}{\includegraphics[angle=0]{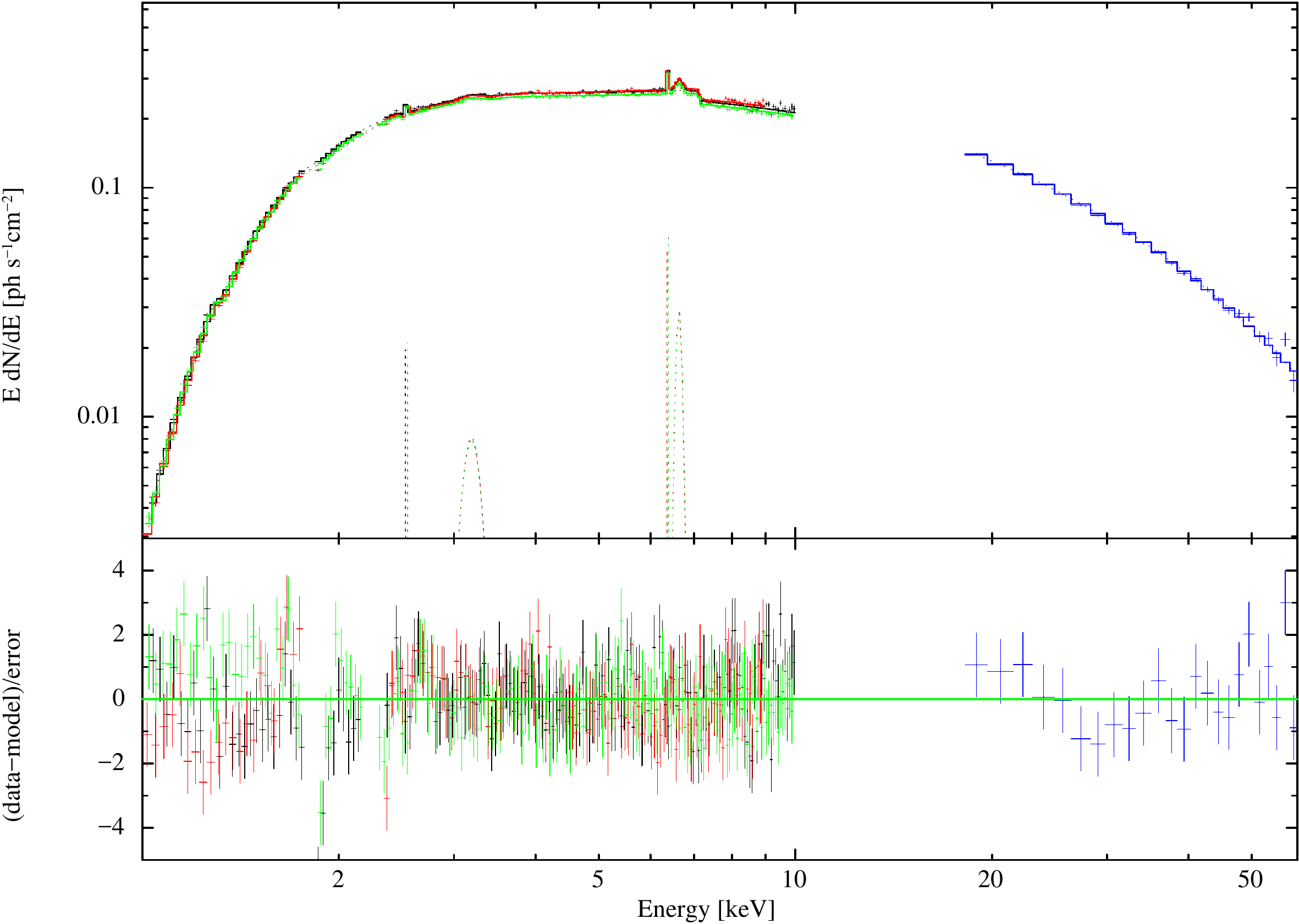}}
\caption{Same as Fig.~\ref{fig:phase_averaged} but for the \suzaku\ data collected in 2007. 
XIS0 (black), XIS1 (red), XIS3 (green), HXD-PIN (blue), and HXD-GSO (cyan) data have been 
rebinned for plotting purposes.}
\label{fig:phase_averaged_2007}
\end{center}
\end{figure}

\section{Phase-resolved spectroscopy}
\label{sec:phase}

To perform the phase-resolved spectral analysis of the \xmm observation we divided the source spin period in 128 phase bins 
and extracted the corresponding \pn and MOS1 spectra. 
We restricted the analysis of the 2012 and 2007 \suzaku\ observations to 20 phase bins due to the limited timing resolution 
of these data. The corresponding XIS, PIN, and GSO spectra were then extracted. All spectra were grouped
with at least 20 photons per energy bin in addition to the 
optimal oversampling of the energy response to guarantee the usability of the $\chi^2$ statistical test.
In all cases, a randomization of the arrival time of each photon within the instrumental time bin has been 
applied to avoid any discretization issue. We fit all spectra  
with the same model employed in Sect.~\ref{sec:spectral}, but freezing few parameters that could not be constrained 
in these lower statistic spectra to the values measured from the phase-averaged analysis.
In particular, we fixed the Galactic column density, as well as the energy 
and widths of all iron lines (these are not expected to change as a function of the NS spin period).  
We also removed from the fit the iron K$\beta$ line, as this feature could not be detected in the phase-resolved spectra. 

Performing fits to all data with the above assumptions still resulted in a number of poorly constrained 
spectral parameters. 
In all phase-resolved \xmm spectra, we fixed the cutoff and folding energies to the average values. Only in spectra, for which
the source flux was low, the column density of the partially covering component turned out to be virtually 
unconstrained and was thus fixed to its average value. 

\begin{figure}[]
\begin{center}
\resizebox{\hsize}{!}{\includegraphics[angle=0]{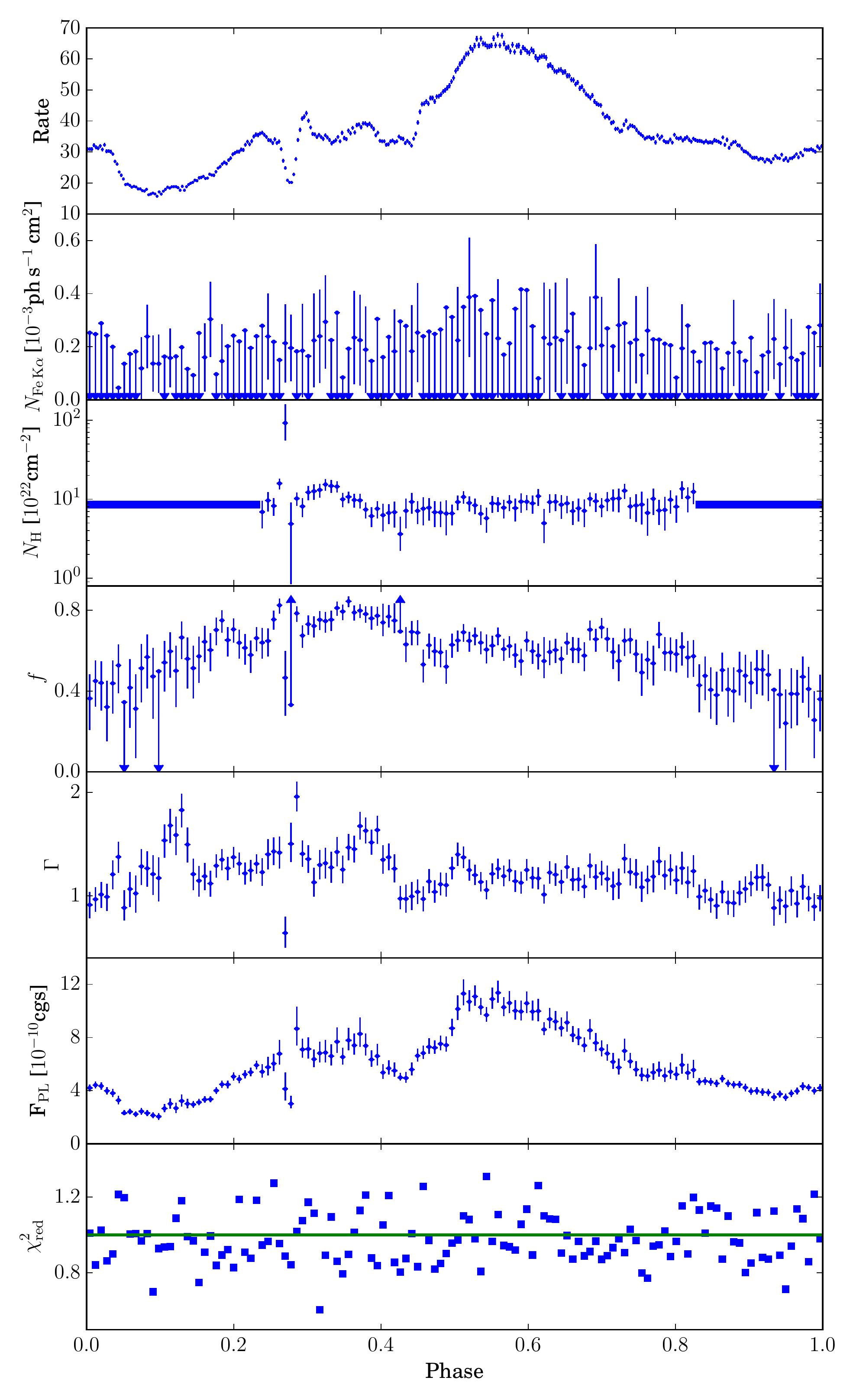}}
\caption{Phase-resolved spectral parameters obtained from the fits to the \xmm data. 
The pulse profile is extracted in 350 bins, while the spectra are extracted in 128 bins (0.5--10\,keV). 
Filled squares in the parameter panels indicate a frozen parameter value. The reduced $\chi^2$ 
is computed with a number of degrees of freedom comprised between 99 and 199, depending on the statistics of the different spectra. 
Uncertainties are reported at 90\% c.l. for all parameters.}
\label{fig:phase_resolved_xmm}
\end{center}
\end{figure}

\begin{figure}[]
\begin{center}
\resizebox{\hsize}{!}{\includegraphics[angle=0]{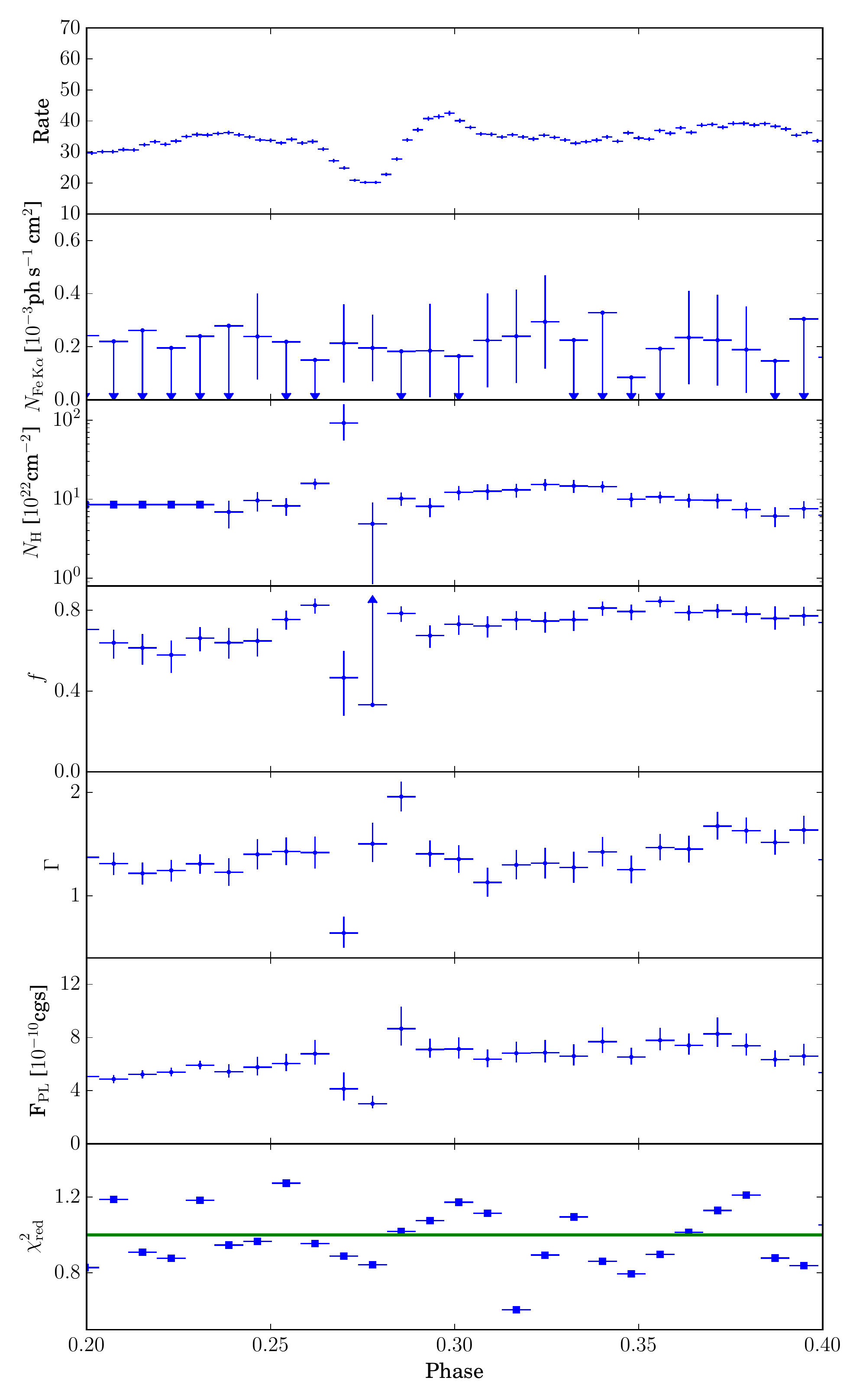}}
\caption{Zoom into the phase-resolved spectral analysis of Fig~\ref{fig:phase_resolved_xmm} around the 
sharp V-shaped feature described in Sect.~\ref{sec:spectral}.}
\label{fig:phase_resolved_zoom}
\end{center}
\end{figure}

\begin{figure}[]
\begin{center}
\resizebox{\hsize}{!}{\includegraphics[angle=0]{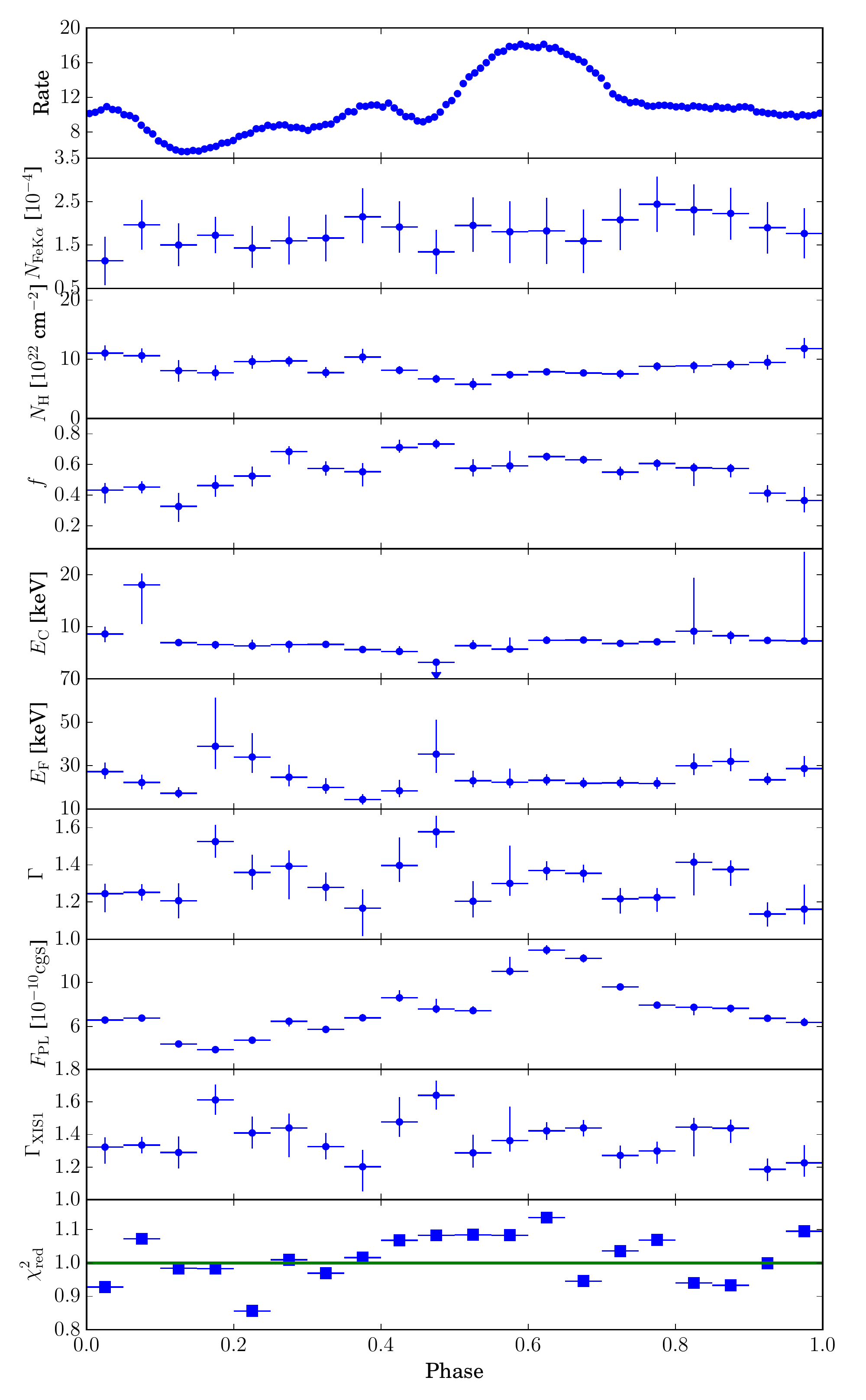}}
\caption{Phase-resolved spectral parameters obtained from the \suzaku\ observation carried out in 2012. 
The pulse profile is extracted in the 0.5--10\,keV energy range using 128 phase bins (a randomization of the photon 
arrival times within the time-resolution unit has been applied in all cases).
Twenty phase-resolved spectra have been extracted due to the lower timing resolution of the \suzaku data 
compared to the \xmm\ ones. 
The normalizations of the iron line are expressed in units of $10^{-4}\,\mathrm{ph\,s^{-1}\,cm^{-2}}$. 
The reduced $\chi^2$ is computed with a number of degrees of freedom comprised between 323
and 378, depending on the statistics of the different spectra.}
\label{fig:phase_resolved_suzaku_12}
\end{center}
\end{figure}

All results of the phase-resolved spectral analysis are summarized in Figs.~\ref{fig:phase_resolved_xmm}--\ref{fig:phase_resolved_suzaku_12}.
The sharp feature described in 
Sect.~\ref{sec:spectral} cannot be detected in the \suzaku\ pulse profiles due to the lower timing resolution. The analysis of the \xmm\ data 
revealed significant spectral changes corresponding to this feature. 
Our analysis suggests that there is a remarkable 
increase in the absorption column density during the ingress into the feature, combined with a strong hardening 
of the power-law slope. After the ingress, the absorption column density decreases and the spectra soften. The partial covering 
parameters corresponding to the out of feature values are recovered shortly before the egress from the feature.
The other spectral variations observed along the pulse profile of the source are in reasonable agreement 
with those measured from the 2012 \suzaku\ data. The difference in the spectral 
parameters measured between the 2007 and 2012 \suzaku\ data were already discussed by \citet{Naik2014}. 
We briefly comment on this point in Sect.~\ref{sec:discussion}.  


\begin{figure}[]
\begin{center}
\resizebox{\hsize}{!}{\includegraphics[angle=0]{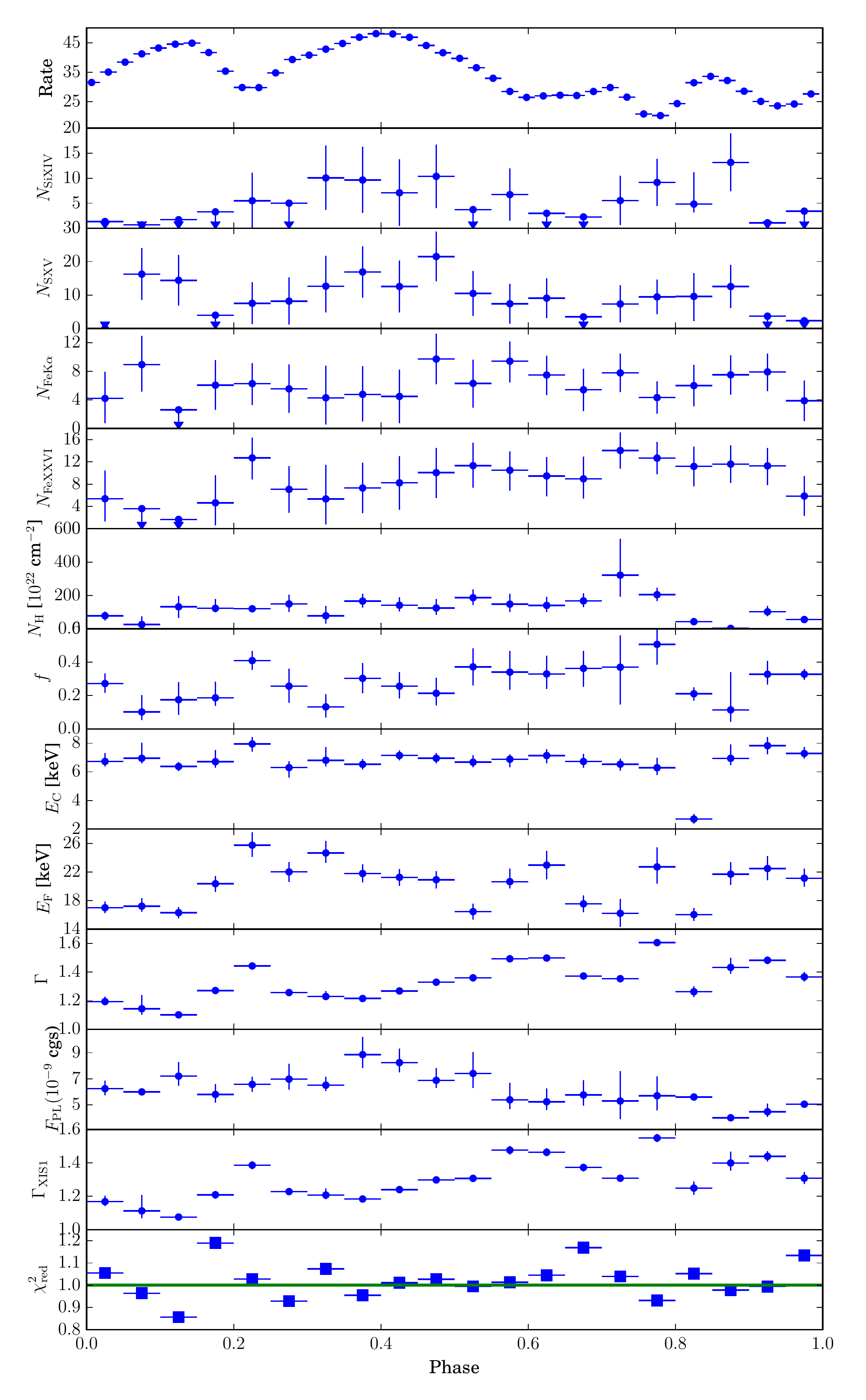}}
\caption{Same as Fig.~\ref{fig:phase_resolved_suzaku_12} but for the \suzaku\ observation carried out in 2007. 
The reduced $\chi^2$ is computed with a number of degrees of freedom comprised between 402 and 435, 
depending on the statistics of the different spectra.}
\label{fig:phase_resolved_suzaku_07}
\end{center}
\end{figure}

\section{Discussion}
\label{sec:discussion}

We reported on the first \xmm\ observation of \exo\ performed in 2014 during the rise to one of the source 
type-I X-ray outbursts. 
The large collecting areas and good timing resolution of the EPIC cameras on-board \xmm\ 
allowed us to study the source pulse profile with high accuracy and measure 
spectral energy variations as a function of the source spin phase.  

The spectral energy distribution of \exo\ in the \xmm\ energy band could be well described by using 
a phenomenological model comprising a power-law with an exponential cutoff at high energies 
and two absorption components. One of these takes into account the presence of the Galactic and circum-binary obscuring material 
in the direction of the X-ray source \citep[note that the expected Galactic column density in the direction of 
\exo\ is $\sim0.9\times 10^{22}\,\mathrm{cm^{-2}}$, ][]{Kalberla2005}, while the second is introduced to describe the effect of a spin-dependent 
partially covering medium in the closest proximities of the NS.
Even though the latter is expected to be ionized due to the relatively high impinging X-ray flux from the 
compact object, we were unable to reveal any spectral signature in the \xmm\ spectra that could be related to the 
ionization, as revealed by the tight upper limit on the ionizations parameter, obtained using the  
\texttt{warmabs} model. This is somehow unexpected based on simple calculations. If we consider the source intrinsic X-ray 
luminosity measured by \xmm,\ $L_X\sim1.1\times10^{37}\,\mathrm{erg/s}$ (derived from the unabsorbed 
0.1--100 keV flux $\sim1.8\times10^{-9}$\,erg/s/cm$^{2}$ at 7.1\,kpc), we can estimate the average mass accretion rate as  
$\dot m = 5.8\times10^{16}\,\mathrm{g/s}$ (assuming that all the potential energy of the accreting matter 
is converted into X-ray radiation). For a NS with a mass (radius) of 1.4\,M$_\odot$ (10\,km) and a  
dipole magnetic field of $5.4\times10^{12}$\,G \citep[as in \exo; see, e.g.,][]{klochkov2007}, the 
accretion column is expected to have a radius of $\sim$300\,m on the star surface \citep[we considered that the 
accretion flow is uniform within the accretion column and that the latter is intercepting the 
inner boundary of the accretion disk at the Alfv\'en radius given by Eq.\,6.19 of][]{frank02}. 
In the free-fall approximation, 
the particle density is thus expected to be of the order of $10^{20}\,\mathrm{cm^{-3}}$ close to the NS surface and to decrease   
by a factor of two 3\,km above (reaching a minimum of $\sim 10^{12}\,\mathrm{cm^{-3}}$ at the Alfv\'en radius). 
As the ionization parameter $\xi=\frac{L_X}{nd^2}$ 
\citep[where $d$ is the distance of the medium from the illuminating source,][]{kallman04} 
turns out to be larger than $10^5$ up to the Alfv\'en radius, the bulk of the accreting 
material around the NS magnetic field lines is expected to be highly ionized. We note that this issue is also observed, e.g.,
in the case of the similar system KS~1947+300 \citep{Ballhausen2016}.

A way to solve this apparent issue is to assume that either the bulk of the X-ray radiation from the NS is highly beamed and it is not 
(always) sufficiently illuminating the accretion stream, or the material in the accretion stream is inhomogeneous and 
over dense compared to our previous estimate. The assumption of an inhomogeneous and over-dense accretion flow seems to 
be in agreement with the low covering fraction measured by our spectral analysis (see Table~\ref{tab:phase_averaged}, Fig.~\ref{fig:phase_resolved_zoom}, 
and Fig.~\ref{fig:phase_resolved_suzaku_12}) and it is known to be more likely to occur in presence of complex accretion geometries,
owing to the tilt between the magnetic and rotational axes and the coupling between the NS magnetosphere 
with the accretion disk \citep[see, e.g.,][]{Meszaros1984}. As an example, hollow accretion columns would naturally lead to higher densities. 
It has been also suggested in the literature that the typical variability of \exo\ and other 
high mass X-ray binaries could be related to the presence of a clumpy (and thus strongly inhomogeneous) stream being accreted 
onto the NS \citep{klochkov2011}. A beamed radiation would certainly not be surprising in the case of 
a young X-ray pulsar due the well known angular dependence of the scattering cross section 
in presence of a strong magnetic field. The latter can lead to the X-ray radiation being preferentially emitted along the direction of the 
stream at low luminosities (pencil beam) or perpendicularly with respect to the stream (fan beam) at higher luminosities. The second possibility 
would be favorable to reduce the ionization state of the accreting material, but we note that a combination of simultaneous fan and 
pencil beams could also be a possibility \citep[see, e.g.,][]{Leahy2004}. 

By analyzing two \suzaku\ observation of \exo\ during two type-I outbursts occurred in 2007 and 2012, we showed that the spectral 
model used to fit the phase averaged \xmm\ data is also suitable to describe the X-ray emission over a 
broader energy range (1-100\,keV). 
Although the 2012 \suzaku\ observation caught the source at a similar luminosity as the one 
recorded from the \xmm\ data, a number of differences in the spectral parameters were measured 
during the analysis carried in Sect.~\ref{sec:spectral} and in Sect.~\ref{sec:phase}. 
Due to inter-calibration uncertainties between the \xmm\ and \suzaku\ instrument we could 
not rule out that some of these discrepancies were instrumental, but it is also 
possible   
that some of the differences in spectral parameters could be the result  
of the accretion process different properties and physical 
conditions in the accretion columns (the two outbursts are separated by 
many orbital revolutions). 
A more reliable comparison of the source high energy spectral distribution in different outbursts 
can be carried out between the results obtained from the 
fits performed on the \suzaku\ data collected in 2012 and 2007. By looking at Table~\ref{tab:phase_averaged}, we note  
a dramatic increase of the column density of the partial covering component with luminosity
and an opposite behavior for the covering fraction  
\citep[as also noticed by][]{Naik2012,Naik2014}. 
The fluorescence lines corresponding to higher ionization states of the iron ions and by Sulfur and Silicon
become detectable in the higher luminosity observation due to the increased  
ionizing X-ray flux. From Fig.~\ref{fig:phase_resolved_suzaku_07}, 
it can be noticed that the neutral iron emission line intensity is roughly constant
along the pulse phase for all datasets, while the lines of the higher ionization stages show significant changes,
which are suggestive of variable illuminations effects. This would be consistent with the idea of an inhomogeneous medium 
and an asymmetric radiation beam.
\begin{figure}[]
\begin{center}
\resizebox{\hsize}{!}{\includegraphics[angle=0]{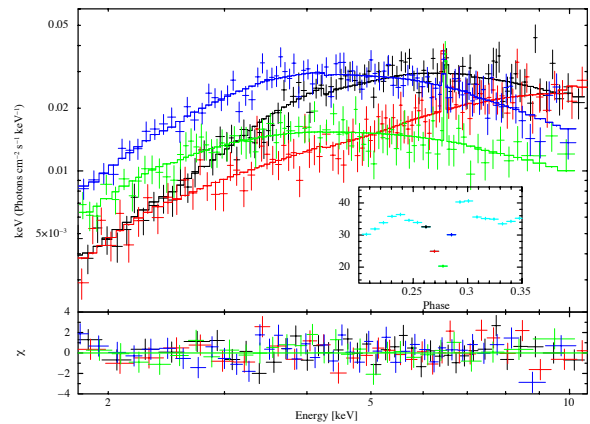}}
\caption{Upper panel: phase-resolved deconvolved energy spectra extracted from the \xmm observation during the V-shaped feature
analyzed in Sect.~\ref{sec:phase}.  The best fit model for all spectra is the one described in Sect.~\ref{sec:spectral}.
The different selected phases for which the spectra are extracted are represented with the same color in the
zoomed pulse profile in the inset (cyan points represent the pulse profile data for which no spectrum is displayed). 
Only \pn data have been shown for clarity. The lower panel shows the residuals from the fit.}
\label{fig:compare_4}
\end{center}
\end{figure}

From the phase-resolved spectral analysis of the \xmm\ observation, we studied with unprecedented detail
spectral changes of the source emission in the soft X-ray energy band as a function of the spin phase. 
The overall behavior of the main parameters is relatively similar in the \xmm\ and the \suzaku\ data 
collected at comparable luminosities. 
In the \pn data, we revealed a remarkable spectral variation during the the ingress and egress of 
a sharp V-shaped feature profile, which was never detected before  (see also Fig.~\ref{fig:compare_4}). Note that a different dip-like 
structure in the pulse profile of \exo\ was also reported by \citet{klochkov2008} using \inte\ data collected around the maximum 
of the source giant outburst occurred in 2006. This feature appeared only above 10\,keV and vanished above 70\,keV. Therefore, 
it could have a completely different origin compared to that detected by \xmm\ in the soft X-rays.
According to the fits, this spectral variability can be 
ascribed mostly to the strong increase of the column density of the ionized absorber during the ingress into 
the feature, and to the hardening of the spectrum (changes in the power-law photon index $\Gamma$ from $\sim$0.7 to 2 are observed 
across the V-shaped feature). The source seems then to recover the usual spectral parameters after the egress from the feature. 
As the density of the accretion stream near the NS is $\sim10^{20}\,\mathrm{cm}^{-3}$, variations of the order of 
$10^{22} \,\mathrm{cm}^{-2}$ in the column density can be expected if the line of sight to the observer in the direction of the 
X-ray source is obscured by just a few meters of stratified accreting material. This part of the accretion column 
can also become a source of reprocessed radiation, which intensity and hardness can change depending on the viewing angle in different 
the phase-resolved spectra. We thus interpret the 
V-shaped feature as being due to the obscuration of the compact emitting region on the NS surface by the accretion column 
that passes in front of the observer line of sight and produces 
an enhanced scattered radiation at the egress. 
As the obscuration occurs on a time scale of $\sim$1/128 $P_\mathrm{spin}\sim 0.3$\,s, which 
corresponds to a linear scale of $\sim$500\,m onto the NS surface, the \xmm\ observation provides also 
constraints on the lateral extension of the accretion column in \exo.\ 
According to this interpretation, we would thus be directly glancing through 
the NS accretion column during the small time interval corresponding to the V-shaped feature in the source pulsed profile. 

The presence of a V-shaped feature could not be verified in the 2012 \suzaku\ data due to their limited timing resolution. 
At odds with \citet{Naik2014}, our phase-resolved spectral analysis did not reveal any significant increase 
in the absorption column density at phase 0.7--0.9, corresponding to the second peak in the hard X-ray pulse profile. 
We also inspected the 2007 \suzaku\ data for the presence of absorption-driven structures in the pulse profile. 
In Fig.~\ref{fig:phase_resolved_suzaku_07}, we noticed that there is a relatively broad dip structure at phase 0.75, which 
could be the high-luminosity counterpart of the dip observed in the \xmm data set. In this case, we should also 
assume that the peak at phase 0.15 in the 2007 \suzaku data corresponds to the main peak of the source pulse profile observed by \xmm\  
at phase 0.6. The dip structure in the 2007 \suzaku\ data spans $\sim1/20$ of the spin phase, translating into a projected scale on 
the NS surface of about 3\,km. Based on a simple scaling relation of the NS hot-spot angular size on the NS surface $\theta$, as function of the X-ray 
luminosity \citep[$\theta \propto L_X^{1/7}$,][]{lamb1973}, we would expect an increase of only about 40\% in the lateral 
size of the NS accretion column between the \xmm\ and the 2007 \suzaku\ observations. A more likely possibility is 
that the dip structure in the 2007 \suzaku\ data is due to the obscuration effect of a larger section of the accretion stream 
farther above the NS surface. Assuming a dipolar field, an increase in the angular size of the absorber by a factor of 6 
would imply that the obscuring region is located at about 40 stellar radii from the NS surface (without accounting for any relativistic effect). 
The results displayed in Fig.~\ref{fig:phase_resolved_suzaku_07} show a marginal indication of an increase in the local 
absorption column density and covering fraction at the phase corresponding to the 
dip, further supporting this interpretation.

\section{Conclusion}
\label{sec:conclusion}

The discovery of a sharp and narrow dip-like feature in the soft X-ray pulse profile of \exo\ during a  
typical type-I X-ray outburst allowed us to have a novel insight on the physical properties of 
the accretion flow in this object. The presence of such feature is so far unique among all known high mass X-ray binaries hosting 
strongly magnetized stars (at the best of our knowledge). Further investigations on other similar systems with X-ray instruments 
endowed with a good timing resolution and large effective areas at $\sim$10\,keV could complement the existing legacy data, 
typically available only at higher luminosities and harder X-rays, and allow us to probe in more details 
the accretion geometry and magnetic field topology of these systems. 

A model comprising a phenomenological Comptonization continuum and a combination of homogeneous and inhomogeneous
absorbers is shown to provide a reasonably 
good fit to the \xmm\ data of \exo,\ as well as to the broader energy coverage spectra of the source provided by the \suzaku observations. 
Exploiting the suitability of such model 
to fit the X-ray spectra of other BeXRBs will permit to possibly achieve a more homogeneous description of the high-energy 
emission from these sources. 

\begin{acknowledgements}
This work is based on observations obtained with \xmm  (OBSID 74524), an ESA science mission with instruments and contributions directly funded by ESA Member States and the USA (NASA) and of data obtained from the \suzaku satellite (OBSIDs 402068010 and 407089010), a collaborative mission between the space agencies of Japan (JAXA) and the USA (NASA).
PP and AAZ have been supported in part by the Polish
NCN grants 2012/04/M/ST9/00780 and 2013/10/M/ST9/00729. 
LD is supported by the Bundesministerium 
f\"ur Wirtschaft und technologie through the Deutches Zentrum f\"ur Luft und Raumfahrt (grant FKZ~50~OG~1602).
\end{acknowledgements}

\bibliographystyle{aa}
\bibliography{exo_prop}

\end{document}